\documentclass[preprint, nopreprintline]{elsarticle}
\usepackage{fullpage,epsf,amssymb,amsthm,amsfonts,amsmath,latexsym, graphicx,array,extarrows,mathtools,mathstyle}
\usepackage[font=footnotesize,labelfont=bf]{caption}
\usepackage[normalem]{ulem}

\newcommand{\overbar}[1]{\mkern 1.5mu\overline{\mkern-1.5mu#1\mkern-1.5mu}\mkern 1.5mu}

\let\oldsqrt\sqrt
\def\sqrt{\mathpalette\DHLhksqrt}
\def\DHLhksqrt#1#2{%
\setbox0=\hbox{$#1\oldsqrt{#2\,}$}\dimen0=\ht0
\advance\dimen0-0.2\ht0
\setbox2=\hbox{\vrule height\ht0 depth -\dimen0}
{\box0\lower0.4pt\box2}}

\numberwithin{equation}{section}

\begin{document}
\begin{frontmatter}

\title{Boundary terms in quantum field theory and the spin structure of QCD} 
\author[PL]{Peter Lowdon}    
\ead{lowdon@physik.uzh.ch}
\address[PL]{Physik-Institut, Universit\"at Z\"urich, Winterthurerstrasse 190, 8057 Z\"urich, Switzerland}

\begin{abstract}
Determining how boundary terms behave in a quantum field theory (QFT) is crucial for understanding the dynamics of the theory. Nevertheless, boundary terms are often neglected using classical-type arguments which are no longer justified in the full quantum theory. In this paper we address this problem by establishing a necessary and sufficient condition for arbitrary spatial boundary terms to vanish in a general QFT. As an application of this condition we examine the issue of whether the angular momentum operator in Quantum Chromodynamics (QCD) has a physically meaningful quark-gluon decomposition. Using this condition it appears as though this is not the case, and that it is in fact the non-perturbative QCD structure which prevents the possibility of such a decomposition.  
\end{abstract} 

\begin{keyword}
QFT boundary terms \sep QCD \sep proton spin decomposition 
\PACS  03.70.+k \sep 11.10.Cd \sep 12.38.Aw \sep 14.20.Dh  \\
\textit{Report number:} ZU--TH 26/14
\end{keyword}
\end{frontmatter}

\setcounter{page}{1}
\pagestyle{plain}

\newtheorem{mydef}{Definition}
\newtheorem{theorem}{Theorem}
\newtheorem{lemma}{Lemma}
\newtheorem{corr}{Corollary}
\newcommand{\lpipe}{\rule[-0.4ex]{0.90pt}{2.3ex}}
\setlength\fboxrule{0.5mm}
\setlength\fboxsep{4mm}

\section{Introduction}
\label{section1}

When classical and quantum field theories are discussed it is often assumed that spatial boundary terms do not contribute~\cite{weinberg_1}-\cite{srednicki}. The standard reasoning given for this is that the dynamical fields in the theory vanish at spatial infinity. Although there may be instances in classical field theory where this boundary condition is applicable, generally this condition is too restrictive. Plane waves~\cite{leader_review}, or cases where the space of field configurations has a non-trivial topology~\cite{weinberg_2}, are two such examples where field solutions may not vanish asymptotically. In the quantum case, axiomatic formulations of field theory assert that fields $\varphi$ are \textit{operator-valued distributions}~\cite{wightman}. Distributions are continuous linear functionals which map a space of \textit{test functions} $\mathcal{T}$ onto the complex numbers: $\varphi: \mathcal{T} \rightarrow \mathbb{C}$. In quantum field theory (QFT), $\mathcal{T}$ is chosen to be some set of space-time functions; usually either the space of continuous functions with compact support $\mathcal{D}(\mathbb{R}^{1,3})$, or the space of Schwartz functions\footnote{The distributions with which these test functions are smeared are called \textit{tempered distributions}.} $\mathcal{S}(\mathbb{R}^{1,3})$. In either of these cases one can represent the image of the map $\varphi$ on a space-time function $f$ as:
\begin{align}
\varphi(f) = (\varphi,f) : = \int d^{4}x \ \varphi(x)f(x)
\label{int_rep}
\end{align}      
which gives meaning to the $x$-dependent field expression $\varphi(x)$. Since $\varphi$ is an \textit{operator}-valued distribution in QFT, only the smeared expression $\varphi(f)$ is guaranteed to correspond to a well-defined operator. The derivative of a distribution $\varphi'$ is defined by:
\begin{align}
(\varphi',f) := -(\varphi,f')
\end{align} 
and is itself also a distribution~\cite{wightman}. By applying the integral representation in equation~\ref{int_rep}, one can interpret this definition as an integration by parts where the boundary terms have been `dropped':
\begin{align}
\int d^{4}x \ \varphi'(x)f(x) = -\int d^{4}x \ \varphi(x)f'(x)
\end{align}   
Although this shorthand notation is useful, and will be used for the calculations in this paper, it can also be slightly misleading. Sometimes it is incorrectly stated that integration by parts of quantum fields can be performed, and the boundary terms neglected. However, distributions are generally not point-wise defined, so boundary expressions like: $\int_{\partial \mathbb{R}^{3} } \varphi(x)f(x)$ are often ill-defined. Therefore, when manipulations like this are performed one is really just applying the definition of the derivative of a distribution, there are no boundary contributions. This makes the question of whether spatial boundary term operators vanish a more subtle issue in QFT than in the classical case. \\

\noindent
The physical rationale behind using operator-valued distributions as opposed to operator-valued functions in QFT is because operators inherently imply a measurement, and this is not well-defined at a single (space-time) point since this would require an infinite amount of energy~\cite{haag}. Instead, one can perform a measurement over a space-time region $\mathcal{U}$, and model the corresponding operator $\mathcal{A}(f)$ as a distribution $\mathcal{A}$ smeared with some test function $f$ which has support in $\mathcal{U}$. If one were to smear $\mathcal{A}$ with another test function $g$, which has different support to $f$, then in general the operators $\mathcal{A}(f)$ and $\mathcal{A}(g)$ would be different. But the interpretation is that these operators measure the \textit{same} quantity, just within the different space-time regions: $\text{supp}(f)$ and $\text{supp}(g)$. \\ 

As well as differentiation it is also possible to extend the notion of \textit{multiplication by a function} to distributions. Given a distribution $\varphi$, a test function $f$, and some function $g$, this is defined as:
\begin{align}
(g\varphi,f) := (\varphi,gf)
\label{dist_prop}
\end{align}   
In order that $g\varphi$ defines a distribution in the case where $f\in \mathcal{D}$, it suffices that $g$ be an infinitely differentiable function. For tempered distributions, in which $f\in \mathcal{S}$, it is also necessary that $g$ and all of its derivatives are bounded by polynomials~\cite{wightman}. \\

\noindent
Besides the assumption that fields are operator-valued distributions, axiomatic approaches to QFT usually postulate several additional conditions that the theory must satisfy. Although different axiomatic schemes have been proposed, these schemes generally contain a common core set of axioms\footnote{See~\cite{wightman},\cite{haag} and~\cite{ojima_book} for a more in-depth discussion of these axioms and their physical motivation.}. For the purpose of the calculations in this paper, the core axioms which play a direct role are:  

\vspace{3mm}

\begin{enumerate}

\item {\bf{Local (anti-)commutativity}}
\begin{align*}
&\textit{If the support of the test functions} \hspace{2mm} f,g, \hspace{1mm} \textit{of the fields} \hspace{2mm} \Psi,\Phi, \hspace{1mm} \textit{are space-like separated then:} \\
&\hspace{35mm} [\Psi(f),\Phi(g)]_{\pm} =\Psi(f)\Phi(g)\pm \Phi(g)\Psi(f) = 0 \\
&\textit{holds when applied to any state vector, for any fields} \hspace{2mm} \Psi,\Phi
\end{align*} 

\item {\bf{Non-degeneracy}}
\begin{align*}
&\textit{The inner product} \hspace{1mm} \langle \cdot | \cdot \rangle  \hspace{1mm} \textit{on the space of states} \hspace{2mm} \mathcal{V} \hspace{1mm} \textit{is non-degenerate:} \\
&\hspace{20mm} \langle \Psi | \omega \rangle = 0, \hspace{3mm} \forall |\Psi \rangle \in \mathcal{V} \hspace{5mm} \Longrightarrow \hspace{5mm} |\omega \rangle =0
\end{align*} 

\end{enumerate}

\noindent
The physical motivation for the local (anti-)commutativity axiom is that it imposes a causality restriction on the theory. Since the action of field operators $\Psi(f)$ on states can be interpreted as the performance of a particular measurement in the space-time region $\text{supp}(f)$~\cite{haag}, if another measurement $\Phi(g)$ is performed in a space-time region $\text{supp}(g)$ which is space-like separated to $\text{supp}(f)$, the axiom states that these two measurements must either commute or anti-commute with one another. In physical terms this means that measurements which are performed a space-like distance apart cannot be causally related to one another. A space of fields $\mathcal{F}$ which satisfies this property is called a \textit{local field algebra}~\cite{strocchi}. By contrast, the non-degeneracy axiom can be imposed without any real loss of generality since any vector $|\omega \rangle$ whose inner product with any other state vanishes will not introduce physical effects into the theory that are describable using the inner product~\cite{ojima_book}. Such states $|\omega \rangle$ are therefore physically trivial with regards to the quantum theory, and can hence be set to zero. \\

\noindent
Additional complications arise when defining quantised gauge field theories. This is because the restriction of a theory to be invariant under a gauge group symmetry $\mathcal{G}$, which corresponds to local invariance under some global symmetry group $G$, leads to a strengthened form of the Noether current conservation condition called the \textit{local Gauss law}~\cite{strocchi}:
\begin{align}
J^{a}_{\mu} = \partial^{\nu}G^{a}_{\mu\nu}, \hspace{6mm} G^{a}_{\mu\nu} = -G^{a}_{\nu\mu}
\label{lgl}
\end{align} 
where $J^{a}_{\mu}$ is the Noether current associated with invariance under the global group $G$. Because of this condition it turns out that there are essentially two quantisation strategies~\cite{strocchi}:
\begin{enumerate}
\item One demands that equation~\ref{lgl} holds as an operator equation, which implies that the algebra of fields $\mathcal{F}$ is no longer local. In particular, if a field transforms non-trivially with respect to the group $G$ (i.e. has a non-zero $G$-charge), the field must be non-local. 
\item One adopts a \textit{local gauge quantisation} in which the local Gauss law is modified. This modification ensures that the field algebra remains local (even for charged fields), but necessitates the introduction of an indefinite inner product on the space of states $\mathcal{V}$, and a condition: $\langle \Psi |J^{a}_{\mu} - \partial^{\nu}G^{a}_{\mu\nu} | \Psi \rangle = 0$ for identifying the physical states $|\Psi\rangle \in \mathcal{V}_{\text{phys}} \subset \mathcal{V}$ (the \textit{weak Gauss law}).   
\end{enumerate}
The advantage with the latter approach is that it allows one keep a local field algebra, and so all of the results from local field theory remain applicable. For the purposes of discussion in this paper we will only consider local quantisations, and in particular we will focus on the local BRST quantisation of Yang-Mills theory. A key feature of BRST quantisation is that a gauge-fixing term $\mathcal{L}_{GF}$ is added to the Lagrangian density $\mathcal{L}$. The modified Lagrangian $\mathcal{L}+\mathcal{L}_{GF}$ is no longer gauge-invariant, but remains invariant under a residual \textit{BRST symmetry} with a conserved charge $Q_{B}$. By defining the physical space of states $\mathcal{V}_{\text{phys}} \subset \mathcal{V}$ to be the states which satisfy the \textit{subsidiary condition}: $Q_{B}\mathcal{V}_{\text{phys}}=0$, this ensures that the weak Gauss law is satisfied and that the field algebra $\mathcal{F}$ is local. The introduction of an indefinite inner product on $\mathcal{V}$ also leads to unphysical \textit{negative} norm states, which are generated by the \textit{Faddeev-Popov ghost} degrees of freedom in $\mathcal{L}_{GF}$. In terms of these extended state spaces, the physical Hilbert space is a quotient space of the form\footnote{The bar denotes the completion of $\mathcal{V}_{\text{phys}}\slash \mathcal{V}_{0}$ to include the limiting states of Cauchy sequences in $\mathcal{V}_{\text{phys}}\slash \mathcal{V}_{0}$.} $\mathcal{H}:=\overbar{\mathcal{V}_{\text{phys}}\slash \mathcal{V}_{0}}$, where $\mathcal{V}_{0}$ contains the zero norm states in $\mathcal{V}$~\cite{ojima_book}.  \\

\noindent
The remainder of this paper is structured as follows; in section~\ref{section2} we apply these general QFT properties in order to establish a necessary and sufficient condition for spatial boundary terms to vanish, in section~\ref{section3} we give a theoretical overview of the angular momentum decomposition problem in QCD and why spatial boundary terms are of particular relevance, and in section~\ref{section4} we apply the results of section~\ref{section2} to this problem. Finally, in section~\ref{section5} we summarise our results and discuss their interpretation.

\section{Spatial boundary terms in QFT}
\label{section2}

The aim of this section will be to demonstrate that the general properties of a quantum field theory, some of which were outlined in section~\ref{section1}, are enough to establish a necessary and sufficient condition for spatial boundary terms of the form: $\int d^{3}x \ \partial_{i}B^{i}$ to vanish in the Hilbert space of physical states $\mathcal{H}$~\cite{ojima_book}, where $B^{i}(x)$ is some arbitrary local\footnote{At least with gauge theories, no generality is lost by imposing locality since a local field algebra can always be assumed by adopting a local gauge quantisation~\cite{strocchi}.} field. However, before fully discussing this condition it is important to first outline the differences between the classical and quantum field theory approaches to conserved (and non-conserved) charges. Classically, charges $Q$ are defined to be the spatial integral of the temporal component of some (not necessarily conserved) current density $j^{\mu}(x)$:
\begin{align}
Q = \int d^{3}x \ j^{0}(x)
\label{Q_class}
\end{align}   
However, in QFT $j^{\mu}(x)$ is typically some product of fields, and is therefore an operator-valued distribution\footnote{Generally the product of distributions is not well-defined, and so one must first introduce a regularisation procedure in order to make sense of such products~\cite{haag}.}. This means that because no smearing with a test function has been performed, the quantised version of the classically motivated definition in equation~\ref{Q_class} will generally \textit{not} correspond to a proper operator in QFT~\cite{strocchi}. But given some space-time test function $f$, a well-defined quantum representation of $Q$ can be written:
\begin{align}
Q = \int d^{4}x \ f(x) j^{0}(x) = j^{0}(f) 
\end{align}   
Following the discussion in section~\ref{section1}, $Q$ is interpreted as acting on the space-time region $\text{supp}(f)$. In order to extend the action of $Q$ to the whole of space, by analogy with the classical case, one can choose the following test function\footnote{This is the standard choice of test function chosen in the literature (\cite{ojima_book}-\cite{kastler}) to define charges.}: $f:=\alpha(x_{0})f_{R}({\bf{x}})$, with $\alpha \in \mathcal{D}(\mathbb{R})$ ($\text{supp}\left(\alpha\right) \subset [-\delta,\delta]$, $\delta >0$) and $f_{R} \in \mathcal{D}(\mathbb{R}^{3})$ where:
\begin{align}
\int dx_{0} \ \alpha(x_{0}) = 1, \hspace{10mm}
   f_{R}({\bf{x}}) = \left\{
     \begin{array}{ll}
       1, & |{\bf{x}}| < R \\
       0, & |{\bf{x}}| > R(1+\varepsilon)
     \end{array}
     \right.
\label{test_f}
\end{align}
with $\varepsilon >0$. Because of the way $f_{R}$ is defined this means that $\partial_{i}f_{R}$ vanishes for $|{\bf{x}}| < R$. Using these test functions one can then construct the following well-defined (localised) charge operator $Q_{R}$:
\begin{align}
Q_{R} = \int d^{4}x \ \alpha(x_{0})f_{R}({\bf{x}}) j^{0}(x) 
\end{align}
The reason why $\alpha$ and $f_{R}$ are chosen to have this form is so that in the special case where the quantised version of the charge is genuinely conserved, this definition is in agreement with the classically-motivated form of $Q$ (equation~\ref{Q_class}) in the limit $R \rightarrow \infty$. \\

Setting: $j^{0}=\partial_{i}B^{i}$ it is clear that spatial boundary terms: $\int d^{3}x \ \partial_{i}B^{i}$ are simply a special class of charges. Therefore, when: $\int d^{3}x \ \partial_{i}B^{i}$ is written in the proceeding discussion (for brevity), this actually corresponds to the smeared expression: $\lim_{R \rightarrow \infty} \int d^{4}x \ \alpha(x_{0})f_{R}({\bf{x}}) \partial_{i}B^{i}(x)$. With this notation in mind, one has the following theorem: 
\ \\

\begin{theorem}
\hspace{20mm} $\int d^{3}x \ \partial_{i}B^{i}$ \hspace{1mm} vanishes in $\mathcal{H}$ \hspace{3mm} $\Longleftrightarrow$ \hspace{3mm} $\int d^{3}x \ \partial_{i}B^{i} | 0 \rangle =0$
\label{theorem1}
\end{theorem}
\vspace{5mm}
\begin{proof}[Proof {\bf{$(\Longleftarrow)$:}}] 

Let $\varphi \in \mathcal{F}(\mathcal{O})$ be some (smeared) local operator\footnote{$\mathcal{F}(\mathcal{O})$ is the polynomial algebra generated by field operators smeared with test functions with compact support in the bounded space-time region $\mathcal{O}$.} and let $\alpha \in \mathcal{D}(\mathbb{R})$ and $f_{R} \in \mathcal{D}(\mathbb{R}^{3})$ be the test functions of compact support defined in equation~\ref{test_f}. Then:
\begin{align*}
\int d^{3}x  \left(\partial_{i}B^{i}\right) \varphi |0\rangle &= \int d^{3}x \ \left[\partial_{i}B^{i}, \varphi\right]_{\pm} |0\rangle \\
&= \lim_{R \rightarrow \infty} \int d^{4}x \ \alpha(x_{0})f_{R}({\bf{x}})\left[\partial_{i}B^{i}(x),\varphi\right]_{\pm}|0\rangle \\
&= - \lim_{R\rightarrow \infty}\int d^{4}x \ \alpha(x_{0})(\partial_{i}f_{R}({\bf{x}}))\left[B^{i}(x),\varphi\right]_{\pm}|0\rangle 
\end{align*}
where in the first line one uses the assumption that: $\int d^{3}x \ \partial_{i}B^{i}|0\rangle=0$ (with $|0\rangle$ the vacuum state), and in the last line the definition for the derivative of a distribution is used. Now because of the way $f_{R}$ is defined it follows that: $\text{supp}\left(\partial_{i}f_{R}\right)=\{x \in\mathbb{R}^{3}; \ R \leq  |{\bf{x}}| \leq R(1+\varepsilon) \}$, and so the support of $\alpha\partial_{i}f_{R}$ will be restricted to the space-time points: $(x_{0},{\bf{x}})$, where $|{\bf{x}}|\geq R$ and $|x_{0}|\leq \delta$. So in the limit\footnote{The existence of this limit is guaranteed by the locality of $B^{i}$ and $\varphi$~\cite{strocchi}.} $R\rightarrow \infty$ the supports of $\alpha\partial_{i}f_{R}$, and the test function for which $\varphi$ is implicitly smeared, will become space-like separated. But by the local (anti-)commutativity axiom this implies that the (anti-)commutator in the last line must vanish exactly, and therefore: $\int d^{3}x  \left(\partial_{i}B^{i}\right) \varphi |0\rangle=0$. The vanishing of the (anti-)commutator is independent of the explicit form for $\alpha$ because $\alpha$ is continuous, has compact support, and is therefore bounded, which means that $\alpha\partial_{i}f_{R}$ will vanish wherever $\partial_{i}f_{R}$ does. Moreover, it follows directly from local (anti-)commutativity that the vanishing of the (anti-)commutator is also independent of the explicit form of $f_{R}$~\cite{kastler}. Because of the vanishing of: $\int d^{3}x  \left(\partial_{i}B^{i}\right) \varphi |0\rangle$, one has:
\begin{align*}
\langle \Psi | \int d^{3}x \ \partial_{i}B^{i} (\varphi |0\rangle)=0, \hspace{3mm} \forall |\Psi\rangle \in \mathcal{H}
\end{align*}
and applying the \textit{Reeh-Schlieder Theorem}\footnote{The Reeh-Schlieder Theorem implies that $\mathcal{H}=\overline{\mathcal{F}(\mathcal{O})|0\rangle}$ for any bounded open set $\mathcal{O}$, where the closure is with respect to some suitable topology. See~\cite{ojima_book} for a proof and in-depth discussion of this theorem.} implies:
\begin{align*}
\langle \Psi | \int d^{3}x \  \partial_{i}B^{i}  | \Phi \rangle =0, \hspace{3mm} \forall |\Psi\rangle,|\Phi\rangle \in \mathcal{H}
\end{align*}
which is precisely the statement that the spatial boundary term: $\int d^{3}x \ \partial_{i}B^{i}$ vanishes in $\mathcal{H}$.
\end{proof}

\begin{proof}[Proof {\bf{$(\Longrightarrow)$:}}] 
Conversely, if one assumes that the spatial boundary term vanishes in $\mathcal{H}$ then:
\begin{align*}
\langle \Psi | \int d^{3}x \ \partial_{i}B^{i} | 0 \rangle =0, \hspace{3mm} \forall |\Psi\rangle \in \mathcal{H}
\end{align*}   
since $| 0 \rangle \in \mathcal{H}$. But this inner product between states is taken to be non-degenerate, so the above condition implies that: $\int d^{3}x \ \partial_{i}B^{i} | 0 \rangle =0$.

\end{proof}
\ \\
\noindent
One subtlety in establishing Theorem~\ref{theorem1} comes from a property which is well-established in Quantum Electrodynamics (QED)~\cite{strocchi_77}, as well as other gauge theories~\cite{ojima_79} -- \textit{charged states are non-local}. This means that it is not possible to create a charged state by applying a \textit{local} operator to the vacuum. However, by virtue of the Reeh-Schlieder Theorem, a charged state can always be approximated by local states as closely as one likes in the sense of convergence in some allowed topology on $\mathcal{H}$. Often this topology is chosen to be the \textit{weak topology}\footnote{Generally it is desirable that the Hilbert space topology be an \textit{admissible topology}, which means that the continuity of a linear functional $\ell$ on $\mathcal{H}$ is equivalent to the existence of a vector $|\Phi\rangle \in \mathcal{H}$ such that: $\ell\left(|\Psi\rangle \right) = \langle \Phi | \Psi \rangle$. The weakest admissible topology is the weak topology~\cite{ojima_book}.} and so convergence means \textit{weak convergence}. Therefore, given that $|\Phi\rangle \in \mathcal{H}$ is a charged state, there exists a sequence of \textit{local} operators $\{\varphi_{n}\}$ such that $\lim_{n \rightarrow \infty} \langle \Psi |\varphi_{n}|0\rangle = \langle \Psi | \Phi \rangle, \ \forall |\Psi\rangle \in \mathcal{H}$. In QED it has in fact been explicitly shown that physical charged states can be constructed from  weak limits of local states~\cite{strocchi_03}. With this convergence property in mind, the proof of Theorem~\ref{theorem1} can then still be shown to hold in the case where $|\Phi\rangle$ is a charged state, since one can apply the same steps as before with $\varphi$ replaced by $\varphi_{n}$, conclude that: $\langle \Psi | \int d^{3}x \ \partial_{i}B^{i} (\varphi_{n} |0\rangle)=0, \ \forall |\Psi\rangle \in \mathcal{H}$, and then take the limit $n \rightarrow \infty$. Moreover, because the (anti-)commutator in the proof is shown to vanish regardless of the explicit form of both $\alpha$ and $f_{R}$, this demonstrates that Theorem~\ref{theorem1} holds independently of the specific test functions in the smearing, and can hence be applied to all spatial boundary term operators of the form: $\int d^{3}x \ \partial_{i}B^{i}$, where $B^{i}(x)$ is any local field.  \\

\noindent
The proof of Theorem~\ref{theorem1} is based on similar discussions in~\cite{ojima_book}-\cite{kastler}, which address the issue of defining a consistent local charge operator and its action on states in $\mathcal{H}$. The surprising conclusion of this theorem is that the vanishing of a spatial boundary term only requires that the corresponding operator annihilates the vacuum state -- \textit{it is independent of how this operator acts on the full space of states.} This result has interesting physical consequences for any QFT, but in particular its relevance to the angular momentum decomposition problem in Quantum Chromodynamics (QCD) will be discussed in sections~\ref{section3} and \ref{section4}. \\  

\noindent
Generally, Theorem~\ref{theorem1} demonstrates that a spatial boundary term operator annihilating the vacuum state is both a necessary and a sufficient condition for that boundary term itself to vanish in the physical Hilbert space $\mathcal{H}$. However, in order to practically determine whether this operator annihilates the vacuum or not, it is easier to instead relate these conditions to equivalent conditions involving matrix elements. This connection is given by the following simple relations:
\begin{align}
 \text{If} \hspace{2mm} \langle \Psi | \int d^{3}x \ \partial_{i}B^{i} |0 \rangle = 0, \ \forall |\Psi\rangle \in \mathcal{H} \hspace{3mm} \Longrightarrow \hspace{3mm}  \int d^{3}x \ \partial_{i}B^{i}|0\rangle = 0   \label{theorem4} \\
 \text{If} \hspace{2mm} \exists  |\Psi\rangle \in \mathcal{H} \hspace{2mm} \text{s.t.} \hspace{2mm} \langle \Psi | \int d^{3}x \ \partial_{i}B^{i} |0 \rangle \neq 0 \hspace{3mm} \Longrightarrow \hspace{3mm}  \int d^{3}x \ \partial_{i}B^{i}|0\rangle \neq 0    \label{theorem5}
\end{align}
Equation~\ref{theorem4} follows immediately from the assumption that the inner product in $\mathcal{H}$ is non-degenerate, and equation~\ref{theorem5} is the logical negation of the statement that the boundary operator acting on the vacuum state is the null vector in $\mathcal{H}$. These relations imply that if one can find \textit{any} state $|\Psi\rangle \in \mathcal{H}$ such that: $\langle \Psi | \int d^{3}x \  \partial_{i}B^{i} |0 \rangle \neq 0$, then this definitively proves: $\int d^{3}x \  \partial_{i}B^{i} |0 \rangle \neq 0$, and hence by Theorem~\ref{theorem1} that: $\int d^{3}x \  \partial_{i}B^{i} \neq 0$. Otherwise, it must be the case that: $\int d^{3}x \ \partial_{i}B^{i} |0 \rangle = 0$, and thus: $\int d^{3}x \ \partial_{i}B^{i} = 0$. To determine the explicit form of these matrix elements one can use the transformation property of fields under translations~\cite{haag}: 
\begin{align}
B^{i}(x) = e^{iP_{\mu}x^{\mu}}B^{i}(0)e^{-iP_{\mu}x^{\mu}}
\end{align}
and insert this between the states $|\Psi\rangle$ and $|0\rangle$ to obtain:
\begin{align*}
\langle \Psi |  \int d^{3}x \ \partial_{i}\Big[ e^{iP_{\mu}x^{\mu}}B^{i}(0)e^{-iP_{\mu}x^{\mu}} \Big] |0 \rangle &= \langle \Psi |  \int d^{3}x \  \Big[ (iP_{i})e^{iP_{\mu}x^{\mu}}B^{i}(0)e^{-iP_{\mu}x^{\mu}} \\
&\hspace{15mm} + e^{iP_{\mu}x^{\mu}}B^{i}(0)e^{-iP_{\mu}x^{\mu}}(-iP_{i}) \Big] |0 \rangle \\
&= \langle \Psi |  \int d^{3}x \ \Big[(iP_{i})e^{iP_{\mu}x^{\mu}}B^{i}(0) \Big] |0 \rangle
\end{align*}
where the second term in the first line vanishes because $P_{\mu}|0\rangle=0$. Now if one takes $|\Psi\rangle$ to be some momentum eigenstate $|p\rangle$, the above relation simplifies to:
\begin{eqnarray}
  \langle p | \int d^{3}x \ \partial_{i}B^{i} |0 \rangle =  \left\{
     \begin{array}{lr}
       0, & \hspace{2mm} p=0 \\
       \lim_{R \rightarrow \infty}  \int d^{4}x \ ip_{i}\alpha(x_{0})f_{R}({\bf{x}}) e^{ip_{\mu}x^{\mu}}\langle p |B^{i}(0) |0 \rangle, & \hspace{2mm} p \neq 0
     \end{array}
     \right.
     \label{cond_1}
\end{eqnarray}

It is interesting to note here that the vacuum expectation value of the spatial boundary operator: $\int d^{3}x \ \partial_{i}B^{i}$ completely vanishes, whereas the off-diagonal matrix element depends on the local term: $\langle p |B^{i}(0) |0 \rangle$.

\section{The proton angular momentum decomposition}
\label{section3}

Theoretical investigations into the spin structure of nucleons have been ongoing ever since the inception of QCD in the 1960s. The evolution of this research area has been influenced by a number of different experimental groups including the European Muon Collaboration (EMC), the Spin Muon Collaboration, and more recently HERMES, COMPASS, STAR and PHENIX~\cite{leader_review}. The focal point of these investigations have largely centred around settling an unresolved dispute known as the \textit{spin crisis}, which refers to results obtained by the EMC experiment~\cite{emc} that suggested quarks accounted for `only' a very small amount of the spin of the proton. Many of the proposed solutions to the spin crisis are based on splitting the QCD angular momentum operator up in different ways, and then arguing a particular physical interpretation of the resulting pieces. It turns out that spatial boundary terms play a prominent role in these decompositions. More specifically, during the rest of this section we will outline why the vanishing of certain spatial boundary \textit{superpotential} terms is an essential assumption in this analysis. \\     

\noindent
As discussed in section~\ref{section1}, in order to analyse the dynamical properties of quantised Yang-Mills theory (in this case QCD), one must choose a quantisation procedure. For the analysis in this paper we will consider the local BRST quantisation with the following Hermitian, gauge-fixed Lagrangian density as proposed by~\cite{ojima_ref1}:
\begin{align}
\mathcal{L}_{QCD} &= -\frac{1}{4}F_{\mu\nu}^{a}F^{\mu\nu a} + \overline{\psi}\left( \frac{i}{2} \gamma^{\mu}(\overrightarrow{\partial}_{\mu}-\overleftarrow{\partial}_{\mu}) + gT^{a}A_{\mu}^{a}\gamma^{\mu} - m \right)\psi + \mathcal{L}_{GF}+ \mathcal{L}_{FP} \label{L_qcd} \\
\mathcal{L}_{GF} &= -(\partial^{\mu}B^{a})A_{\mu}^{a} + \frac{\xi}{2}B^{a}B^{a} \\
\mathcal{L}_{FP} &= -i\partial^{\mu}\overbar{C}^{a}(D_{\mu}C)^{a}
\end{align}   
where $C^{a},\overbar{C}^{a}$ are the Faddeev-Popov ghost fields, $B^{a}$ is the auxiliary gauge fixing field, $\xi$ is a gauge fixing parameter, and one defines: $(D_{\mu}C)^{a}:=\partial_{\mu}C^{a} -gf^{abc}A_{\mu}^{c}C^{b}$. To determine the spin structure of QCD one must first define the energy-momentum tensor of the theory $T_{QCD}^{\mu\nu}$. As with any current density the definition of $T^{\mu\nu}$ is always ambiguous up to a sign, but for the purposes of this paper the following definition will be used:
\begin{align}
T^{\mu\nu} = \frac{2}{\sqrt{-g}} \frac{\delta S}{\delta g_{\mu\nu}}
\end{align}
where: $g=\text{det}g_{\mu\nu}$, $S=\int d^{4}x \sqrt{-g} \mathcal{L}$, and $\mathcal{L}$ has the formal functional dependence: $\mathcal{L}=\mathcal{L}(g_{\mu\nu},\Psi^{I},\nabla_{\alpha}\Psi^{I})$, with $\nabla_{\alpha}$ the general relativistic covariant derivative\footnote{$\nabla_{\alpha}\Psi^{I}$ varies in form depending on the type of field $\Psi^{I}$} and $\{\Psi^{I}\}$ the dynamical fields in the theory (with possible internal index $I$). As was famously shown by Belinfante~\cite{belinfante}, this expression can always be decomposed into the following form:
\begin{align}
T^{\mu\nu} = T^{\mu\nu}_{c} + \frac{1}{2}\partial_{\rho}\left(S^{\rho\mu\nu} + S^{\mu\nu\rho} + S^{\nu\mu\rho}  \right)
\label{belinfante_T}
\end{align} 
where $T^{\mu\nu}_{c}$ is the canonical energy-momentum tensor and $S^{\rho\mu\nu}$ is the so-called spin-angular momentum density term~\cite{ojima_book}:
\begin{align}
T^{\mu\nu}_{c} := \frac{\delta \mathcal{L}}{\delta (\partial_{\mu}\Psi^{I})}\partial^{\nu}\Psi^{I} - g^{\mu\nu}\mathcal{L},
\hspace{10mm} S^{\rho\mu\nu} := -i \frac{\delta \mathcal{L}}{\delta (\partial_{\rho}\Psi^{I})}(s^{\mu\nu})^{I}_{J}\Psi^{J}
\end{align}
$(s^{\mu\nu})^{I}_{J}$ corresponds to the Lorentz generator $s^{\mu\nu}$ in the finite-dimensional representation\footnote{E.g. for a vector field: $(s^{\mu\nu}_{V})^{I}_{J}=i(g^{\mu I}\delta^{\nu}_{\ J}- g^{\nu I}\delta^{\mu}_{\ J})$, where $I,J$ are space-time indices, whereas for a spinor field: $(s^{\mu\nu}_{S})^{I}_{J}= \tfrac{i}{4}[\gamma^{\mu},\gamma^{\nu}]^{I}_{J}$, with $I,J$ spinor indices.} of the field $\Psi^{I}$. It is interesting to note that the question of whether the quantised Belinfante or canonical energy-momentum tensor is physically more relevant is still an on-going issue~\cite{leader_review}. A discussion of the related subtleties between the Belinfante $P^{\mu}$ and canonical $P^{\mu}_{c}$ momentum operators is given in~\ref{appendix_a}. Applying the definition in equation~\ref{belinfante_T} to QCD, one obtains the following expression for the energy-momentum tensor~\cite{ojima_ref1}:
\begin{align}
T^{\mu\nu}_{QCD} &= T^{\mu\nu}_{\text{phys}} - \bigg\{Q_{B}, \ (\partial^{\mu}\overbar{C}^{a})A^{\nu a} + (\partial^{\nu}\overbar{C}^{a})A^{\mu a} + g^{\mu\nu}\left(\frac{1}{2}\xi  \overbar{C}^{a} B^{a} - (\partial^{\rho} \overbar{C}^{a})A_{\rho}^{a} \right) \bigg\} \\
T^{\mu\nu}_{\text{phys}} &= \frac{1}{2}\overline{\psi}\left( \frac{i}{2} \gamma^{\mu}(\overrightarrow{\partial^{\nu}}-\overleftarrow{\partial^{\nu}}) + gT^{a}A^{\nu a}\gamma^{\mu} \right)\psi + \hspace{1mm} (\mu \leftrightarrow \nu) + F^{\mu a}_{\ \rho}F^{\rho \nu a} + \frac{1}{4}g^{\mu\nu}F_{\alpha\beta}^{a}F^{\alpha\beta a} 
\label{T_{qcd}}
\end{align} 
where $Q_{B}$ is the BRST charge. In any field theory the current associated with Lorentz transformations is the rank-3 tensor defined by~\cite{belinfante}:
\begin{align}
M^{\mu\nu\lambda} := x^{\nu}T^{\mu\lambda} - x^{\lambda}T^{\mu\nu} 
\end{align}
Using this definition, the Lorentz current in QCD can be written:
\begin{align*}
M^{\mu\nu\lambda}_{QCD} = x^{\nu}T^{\mu\lambda}_{\text{phys}} - x^{\lambda}T^{\mu\nu}_{\text{phys}} + ix^{[\nu}\delta_{B}\bigg[(\partial^{\mu}\overbar{C}^{a})A^{\lambda] a} + (\partial^{\lambda]}\overbar{C}^{a})A^{\mu a} + g^{\mu\lambda]}\left(\frac{1}{2}\xi  \overbar{C}^{a} B^{a} - (\partial^{\rho} \overbar{C}^{a})A_{\rho}^{a} \right) \bigg] 
\end{align*}
where $\delta_{B}: \mathcal{F} \rightarrow \mathcal{F}$ is the BRST variation map defined for $\mathcal{R} \in \mathcal{F}$ by: 
\begin{align}
\delta_{B}\mathcal{R} := [iQ_{B},\mathcal{R}]_{\pm}
\end{align}  
and $\pm$ signifies an anti-commutator or commutator depending on whether $\mathcal{R}$ has an odd or an even \textit{ghost number}\footnote{The ghost number corresponds to the number of ghost fields contained in a composite operator~\cite{ojima_book}.} respectively. The remarkable thing about this structure is that all of the ghost and gauge fixing fields are contained in a single BRST variation (coboundary) term, and this guarantees that these unphysical operators will not contribute\footnote{Nevertheless, these unphysical fields are still essential for ensuring the consistency and Lorentz covariance of the theory.} to any physical matrix element involving states in $\mathcal{H}$~\cite{ojima_book}. For this reason we will not discuss this term any longer and will simply set: $T^{\mu\nu}_{QCD}\equiv T^{\mu\nu}_{\text{phys}}$. The remaining physical QCD Lorentz current then takes the form:
\begin{align}
M^{\mu\nu\lambda}_{QCD} &= x^{\nu}\Big[ \frac{1}{2}\overline{\psi}\left( \frac{i}{2} \gamma^{\mu}(\overrightarrow{\partial^{\lambda}}-\overleftarrow{\partial^{\lambda}}) + gT^{a}A^{\lambda a}\gamma^{\mu} \right)\psi + \hspace{1mm} (\mu \leftrightarrow \lambda)  + F^{\mu a}_{\ \rho}F^{\rho \lambda a} + \frac{1}{4}g^{\mu\lambda}F_{\alpha\beta}^{a}F^{\alpha\beta a}        \Big] \nonumber \\
&\hspace{5mm} - x^{\lambda}\Big[ \frac{1}{2}\overline{\psi}\left( \frac{i}{2} \gamma^{\mu}(\overrightarrow{\partial^{\nu}}-\overleftarrow{\partial^{\nu}}) + gT^{a}A^{\nu a}\gamma^{\mu} \right)\psi + \hspace{1mm} (\mu \leftrightarrow \nu) + F^{\mu a}_{\ \rho}F^{\rho \nu a} + \frac{1}{4}g^{\mu\nu}F_{\alpha\beta}^{a}F^{\alpha\beta a} 
\Big] \nonumber \\
&= x^{\nu}F^{\mu a}_{\hspace{2mm}\rho}F^{\rho \lambda a} -x^{\lambda}F^{\mu a}_{\hspace{2mm}\rho}F^{\rho \nu a} + \frac{1}{4}F_{\alpha\beta}^{a}F^{\alpha\beta a}(x^{\nu}g^{\mu\lambda}-x^{\lambda}g^{\mu\nu}) \nonumber \\
&\hspace{5mm}+ \frac{i}{4}\Big[x^{\nu}\overline{\psi}( \gamma^{\mu}D^{\lambda} + \gamma^{\lambda}D^{\mu})\psi - (\nu \leftrightarrow \lambda) \Big] + \hspace{1mm} \text{h.c.} 
\label{m_qcd_full}
\end{align}  
where $D^{\mu}:= \partial^{\mu} - igT^{a}A^{\mu a}$ is the covariant derivative and h.c. denotes the Hermitian conjugate. Individually, the terms in equation~\ref{m_qcd_full} do not have a clear interpretation. However, in the early discussion of $M^{\mu\nu\lambda}_{QCD}$ in the literature it was suggested~\cite{Jaffe_Manohar} that a more meaningful expression can be obtained by factoring out total divergence terms. In this case it turns out that one can write~\cite{Jaffe_Manohar}:
\begin{align}
M^{\mu\nu\lambda}_{QCD} = &\frac{i}{2}\overline{\psi}\gamma^{\mu}\left(x^{\nu}\partial^{\lambda} - x^{\lambda}\partial^{\nu} \right)\psi  + \hspace{1mm}\text{h.c.} \hspace{1mm} +\frac{1}{2}\epsilon^{\mu\nu\lambda\rho}\overline{\psi}\gamma_{\rho}\gamma^{5}\psi \nonumber \\
& -F^{\mu\rho a}\left(x^{\nu}\partial^{\lambda} - x^{\lambda}\partial^{\nu}    \right)A_{\rho}^{a}  + F^{\mu\lambda a }A^{\nu a} + F^{\nu\mu a}A^{\lambda a} +\frac{1}{4}F_{\alpha\beta}^{a}F^{\alpha\beta a}\left(x^{\nu}g^{\mu\lambda} -x^{\lambda}g^{\mu\nu} \right) \nonumber \\
&-\frac{i}{16}\partial_{\beta}\Big[x^{\nu}\overline{\psi}\{ \gamma^{\lambda},[\gamma^{\mu},\gamma^{\beta}] \}\psi -(\nu \leftrightarrow \lambda)  \Big] +\partial_{\beta}(x^{\nu}F^{\mu\beta a}A^{\lambda a} - x^{\lambda}F^{\mu\beta a}A^{\nu a})
\label{m_split}
\end{align}
If one then \textit{chooses} to drop these divergence terms, the current takes the following partitioned form: 
\begin{align}
M^{\mu\nu\lambda}_{QCD} =&\frac{i}{2}\overline{\psi}\gamma^{\mu}\left(x^{\nu}\partial^{\lambda} - x^{\lambda}\partial^{\nu} \right)\psi  + \hspace{1mm}\text{h.c} \hspace{1mm} +\frac{1}{2}\epsilon^{\mu\nu\lambda\rho}\overline{\psi}\gamma_{\rho}\gamma^{5}\psi \nonumber \\
& -F^{\mu\rho a}\left(x^{\nu}\partial^{\lambda} - x^{\lambda}\partial^{\nu}    \right)A_{\rho}^{a}  + F^{\mu\lambda a }A^{\nu a} + F^{\nu\mu a}A^{\lambda a} +\frac{1}{4}F_{\alpha\beta}^{a}F^{\alpha\beta a}\left(x^{\nu}g^{\mu\lambda} -x^{\lambda}g^{\mu\nu} \right) 
\label{m_qcd}
\end{align}
This is often referred to in the literature as the \textit{Jaffe-Manohar decomposition}~\cite{leader_review},\cite{Wakamatsu}. The fundamental difference between this expression and the expression in equation~\ref{m_qcd_full} is that the interaction terms between the quark and gluon fields do not feature\footnote{The gluon self-interaction terms do still feature in the expression though.}, and this makes it easier to give the individual remaining terms a physical interpretation. It also turns out that one can arrive at this same decomposition by instead using the canonical energy-momentum tensor $T^{\mu\nu}_{c}$, and in this context this is referred to as the canonical version of the Lorentz current~\cite{leader_review}. In a similar manner to the above derivation, it is also necessary in this case to drop certain divergence terms in order to arrive at the decomposition in equation~\ref{m_qcd}. By applying the definition for the angular momentum charge:
\begin{align}
J^{i}_{QCD}:= \frac{1}{2} \ \epsilon^{ijk}\int d^{3}x \ M^{0jk}_{QCD}(x)
\label{j_qcd}
\end{align}
the decomposition of the current in equation~\ref{m_qcd} becomes a decomposition of charges. These charges are often then individually interpreted as corresponding to physically distinct angular momentum sources~\cite{Ji_initial}-\cite{ji_lorentz}. As well as the Jaffe-Manohar decomposition there also exist many other possible ways of decomposing $J^{i}_{QCD}$~\cite{leader_review},\cite{Ji_initial}. Despite the variety in structure of these different decompositions, it turns out~\cite{leader_review} that they all differ from one another by terms of the form: $\partial_{\beta}B^{[\mu\beta][\nu\lambda]}$, which are called \textit{superpotentials}\footnote{The notation $B^{[\mu\beta][\nu\lambda]}$ implies that $B^{\mu\beta\nu\lambda}$ is anti-symmetric in the indices $(\mu,\beta)$ and $(\nu,\lambda)$.}~\cite{Jaffe_Manohar}. Therefore, in order for any of these decompositions to hold one is required to drop superpotential terms, and if this procedure is justified it implies that all of the possible decompositions are physically equivalent to one another~\cite{leader_review}. \\

\noindent
The argument in the literature~\cite{Jaffe_Manohar} for dropping superpotential terms goes as follows: if two current densities $\widetilde{M}$ and $M$ differ by a superpotential:
\begin{align*}
\widetilde{M}^{\mu\nu\lambda} = M^{\mu\nu\lambda} + \partial_{\beta}B^{[\mu\beta][\nu\lambda]}
\end{align*}
then the corresponding charges $\widetilde{J}^{\nu\lambda}=\int d^{3}x \ \widetilde{M}^{0\nu\lambda}$ and $J^{\nu\lambda}=\int d^{3}x \ M^{0\nu\lambda}$ must be related by:
\begin{align*}
\widetilde{J}^{\nu\lambda} = J^{\nu\lambda} + \int d^{3}x \hspace{1mm} \partial_{i}B^{[0i][\nu\lambda]}
\end{align*}
Assuming that $B^{[0i][\nu\lambda]}(x)$ vanishes at spatial infinity, the divergence theorem implies that: $\widetilde{J}^{\nu\lambda} = J^{\nu\lambda}$, and so for computing charges the currents $\widetilde{M}^{\mu\nu\lambda}$ and $M^{\mu\nu\lambda}$ are indistinguishable. However, the results and discussions in sections~\ref{section1} and~\ref{section2} demonstrate that this argument is often too simplistic in the classical case, and is also not transferable to the corresponding quantised theory. Nevertheless, these arguments for dropping boundary terms have been applied to many situations, one of which being the derivation of the \textit{proton angular momentum sum rule}~\cite{Jaffe_Manohar},\cite{jaffe_delta_g}. This derivation starts by taking the full expression for $M^{\mu\nu\lambda}_{QCD}$ (equation~\ref{m_split}) and inserting it into equation~\ref{j_qcd}, giving:
\begin{align}
J^{i}_{QCD} &= \underbrace{ \epsilon^{ijk}\int d^{3}x \left[ \frac{i}{2}\overline{\psi}\gamma^{0}\left(x^{j}\partial^{k} \right)\psi   + \hspace{1mm}\text{h.c.} \right] }_{:= L_{q}^{i}} \hspace{1mm} 
+ \underbrace{ \epsilon^{ijk}\int d^{3}x \left[ \frac{1}{4}\epsilon^{0jkl}\overline{\psi}\gamma_{l}\gamma^{5}\psi \right]}_{:= S_{q}^{i}} \nonumber \\
& \underbrace{-\epsilon^{ijk}\int d^{3}x \left[ F^{0l a}\left(x^{j}\partial^{k} \right)A_{l}^{a} \right] }_{:=L_{g}^{i}}  + \underbrace{\epsilon^{ijk}\int d^{3}x \left[ F^{0k a }A^{j a} \right] }_{:= S_{g}^{i}}  \nonumber \\
&\underbrace{-\frac{i}{16}\epsilon^{ijk}\int d^{3}x \ \partial_{l}\Big[x^{j}\overline{\psi}\{ \gamma^{k},[\gamma^{0},\gamma^{l}] \}\psi  \Big]}_{:=\mathcal{S}_{1}^{i}} + \underbrace{\epsilon^{ijk}\int d^{3}x \ \partial_{l}(x^{j}F^{0l a}A^{k a} )}_{:=\mathcal{S}_{2}^{i}}
\label{full_decomp}
\end{align} 
If one then assumes that the superpotential boundary terms $\mathcal{S}_{1}^{i}$ and $\mathcal{S}_{2}^{i}$ in this expression vanish, and inserts the remaining $z$-components ($i=3$) between $z$-polarised proton states\footnote{There are of course subtleties~\cite{strocchi} in how to define such asymptotic states, but we will not discuss these here.} $|p,s\rangle$, one obtains the proton angular momentum sum rule~\cite{Jaffe_Manohar},\cite{jaffe_delta_g}:
\begin{align}
\frac{1}{2} = \frac{1}{2}\Sigma + L_{q} + S_{g} + L_{g} 
\label{spin_sum_rule}
\end{align}   
where the $\tfrac{1}{2}$ term on the left-hand side comes from the fact that: $J_{QCD}^{3}|p,s\rangle=\tfrac{1}{2}|p,s\rangle$, and each of the other terms are defined by:  
\begin{align*}
\frac{1}{2}\Sigma = \frac{\langle p,s| S_{q}^{3} |p,s \rangle}{\langle p,s| p,s \rangle}, \hspace{3mm}   S_{g} = \frac{\langle p,s| S_{g}^{3} |p,s \rangle}{\langle p,s| p,s \rangle}, \hspace{3mm} L_{q} = \frac{\langle p,s| L_{q}^{3} |p,s \rangle}{\langle p,s| p,s \rangle}, \hspace{3mm}  L_{g} = \frac{\langle p,s| L_{g}^{3} |p,s \rangle}{\langle p,s| p,s \rangle}
\end{align*}
where a sum over quark flavour is also implicitly assumed in the definitions of $\Sigma$ and $L_{q}$. $\Sigma$/$S_{g}$ are then interpreted as the contributions to the $z$-component of the internal spin of the proton from quarks/gluons\footnote{The interpretation of $\Sigma$ and $S_{g}$ comes from the equality of these terms with the corresponding partonic quantities in the infinite momentum frame and $A^{0}=0$ gauge~\cite{Jaffe_Manohar},\cite{jaffe_delta_g}.}, and $L_{q}$/$L_{g}$ the contributions to the $z$-component of the orbital angular momentum of the proton from the quarks/gluons. It is clear that this derivation requires that the superpotential boundary terms either exactly vanish, or at least vanish when inserted between proton states. In the next section we will use the results of section~\ref{section2} to address whether either of these conditions is actually satisfied in this case.

\section{Superpotential boundary terms in QCD}
\label{section4}

In this section we will focus on addressing the issue of angular momentum decompositions in QCD, and in particular whether the proton angular momentum sum rule holds. As discussed in section~\ref{section3}, to tackle this question it is important to understand superpotential spatial boundary terms of the form: $\int d^{3}x \ \partial_{i} \left( x^{j}B^{k 0 i} \right)$, where $B^{k 0 i}(x)$ is local. Despite the explicit $x$-dependence, Theorem~\ref{theorem1} continues to hold with the replacement: $B^{i} \rightarrow x^{j}B^{k 0 i}$ because the function multiplication property of distributions (equation~\ref{dist_prop}) allows one to re-write the boundary operator in terms of $B^{k 0 i}$ smeared with the test function $x^{j}\alpha\partial_{i}f_{R}$, and so the local (anti-)commutativity argument continues to hold. In section~\ref{section3} it was established that in order for the derivation of the proton angular momentum sum rule to remain valid, it must be the case that either one of the following conditions is satisfied:
\vspace{3mm} 
\begin{enumerate}
\item The superpotential boundary term operators $\mathcal{S}^{i}_{1}$ and $\mathcal{S}^{i}_{2}$ are exactly vanishing. 
\item These operators vanish when inserted between identical $z$-polarised proton states $|p,s\rangle$.
\end{enumerate}
\ \\
Using the conditions in equations~\ref{theorem4} and \ref{theorem5} the first statement can be tested as follows:
\begin{align}
\text{If} \hspace{2mm} \exists|p\rangle \in \mathcal{H} \hspace{2mm} \text{s.t:} \ \langle p| \int d^{3}x \ \partial_{i} \left( x^{j}B^{k 0 i}(x) \right) |0\rangle \neq 0 \hspace{3mm} \Longrightarrow  \hspace{3mm} \int d^{3}x \ \partial_{i} \left( x^{j}B^{k 0 i}(x) \right)  \neq 0
\label{stat}
\end{align}
By performing an analogous calculation to the one at the end of section~\ref{section2}, the matrix elements of the superpotential operator: $\int d^{3}x \ \partial_{i} \left( x^{j}B^{k 0 i}(x) \right)$ can be written in the form: 
\begin{eqnarray}
  \langle p | \int d^{3}x \ \partial_{i}\left(x^{j}B^{k 0 i}(x)\right) |0 \rangle =  \left\{
     \begin{array}{lr}
       \lim_{R \rightarrow \infty}  \int d^{3}x \ f_{R}({\bf{x}}) \langle 0| B^{k 0 j}(0) |0 \rangle , &  p=0 \\
        \lim_{R \rightarrow \infty}  \int d^{4}x \ \alpha(x_{0})f_{R}({\bf{x}})e^{ip_{\mu}x^{\mu}} \big[ \langle p| B^{k 0 j}(0) |0 \rangle \\ 
        \hspace{31mm} + ip_{i}\langle p| x^{j}B^{k 0 i}(0) |0 \rangle \big], &  p \neq 0   
     \end{array}
     \right.
       \label{cond_2}
\end{eqnarray}
where $|p\rangle$ is some momentum eigenstate. In the case of the Jaffe-Manohar angular momentum decomposition discussed in section~\ref{section3}, the superpotential boundary terms $\mathcal{S}^{i}_{1}$ and $\mathcal{S}^{i}_{2}$ are given by: 
\begin{align}
\mathcal{S}^{i}_{1} = -\frac{i}{16}\epsilon^{ijk}\int d^{3}x \ \partial_{l}\left(x^{j}\overline{\psi}\{ \gamma^{k},[\gamma^{0},\gamma^{l}] \}\psi \right)  \hspace{10mm} \mathcal{S}^{i}_{2} = \epsilon^{ijk}\int d^{3}x \ \partial_{l}\left( x^{j}F^{0 l a}A^{k a} \right)
\end{align} 
Choosing $|p\rangle=|0\rangle$, and applying equation~\ref{cond_2}, the matrix elements of these operators can then be written:
\begin{align*}
\langle 0| \mathcal{S}^{i}_{1} |0\rangle 
&=  \lim_{R \rightarrow \infty}  \int d^{3}x \ \frac{1}{4}f_{R}({\bf{x}}) \epsilon^{ijk}\epsilon^{0jk l}\langle 0| \overline{\psi}\gamma^{l}\gamma^{5} \psi |0 \rangle \\
\langle 0|\mathcal{S}^{i}_{2}|0\rangle &=  \lim_{R \rightarrow \infty}  \int d^{3}x \ f_{R}({\bf{x}}) \epsilon^{ijk}\langle 0|F^{0 j a}A^{k a} |0\rangle
\end{align*}

\noindent
In both cases these expressions are non-zero if the vacuum expectation values $\epsilon^{ijk}\epsilon^{0jk l}\langle 0| \overline{\psi}\gamma^{l}\gamma^{5} \psi |0 \rangle$ and $\epsilon^{ijk}\langle 0|F^{0 j a}A^{k a} |0\rangle$ are non-zero. It is important to note here that these expressions are non-trivial because the fields involved are solutions to the full interacting theory, and so their expectation values are non-perturbative objects. Moreover, these combinations of fields do not correspond to conserved currents, so one cannot infer their value based on conservation properties. In order to establish the value of expressions such as these one must either employ a non-perturbative technique (such as lattice gauge theory), or make use of additional symmetries to restrict their form. Calculations such as these have been performed in the literature, and there is evidence to suggest that the first of these vacuum expectation values is non-vanishing~\cite{v-axial_paper}. The second expression though has not to our knowledge been computed\footnote{The similar expression $\langle A^{2}\rangle$ has been computed though, using lattice QCD, and was found to be non-zero~\cite{lattice_A_sq}.}. In the special case where one takes: $|\Psi\rangle=|\pi\rangle$ (the pion state) for the expectation value in the first case, then one can use the hypothesised relation:
\begin{align} 
\langle \pi |\overline{\psi}\gamma^{l}\gamma^{5} \psi |0 \rangle = if_{\pi}p^{l}
\end{align} 
where $f_{\pi} \neq 0$ is the \textit{pion form factor}~\cite{pion_form} and $p^{l}$ is the pion's 3-momentum. Inserting this expression into equation~\ref{cond_2} also gives a non-zero result for the $p\neq 0$ case. Applying the condition in equation~\ref{stat}, these results suggest that $\mathcal{S}^{i}_{1}$ is in general non-vanishing. Therefore, since $\mathcal{S}^{i}_{2}$ does not cancel $\mathcal{S}^{i}_{1}$, this casts doubt on the validity of the angular momentum operator decomposition: $J_{QCD}^{i}= S_{q}^{i}+L_{q}^{i}+S_{g}^{i}+L_{g}^{i}$. \\

\noindent
To compute whether condition 2 is satisfied or not one must calculate the matrix elements of the superpotential operators $\mathcal{S}^{i}_{1}$ and $\mathcal{S}^{i}_{2}$ between the $z$-polarised proton states $|p,z\rangle$. Performing this calculation one obtains:
\begin{align*}
\langle p,s| \mathcal{S}^{i}_{1} |p,s\rangle 
&=  \lim_{R \rightarrow \infty}  \int d^{3}x \ \frac{1}{4}f_{R}({\bf{x}}) \epsilon^{ijk}\epsilon^{0jk l}\langle p,s| \overline{\psi}\gamma^{l}\gamma^{5} \psi |p,s \rangle  \\
\langle p,s|\mathcal{S}^{i}_{2}|p,s\rangle &=  \lim_{R \rightarrow \infty}  \int d^{3}x \ f_{R}({\bf{x}}) \epsilon^{ijk}\langle p,s|F^{0 j a}A^{k a} |p,s\rangle
\end{align*}
Without applying a non-perturbative technique it is unclear whether either of these expressions are vanishing or not. However, by computing the same matrix elements for the operators $S_{q}^{i}$ and $S_{g}^{i}$ in equation~\ref{full_decomp}, it turns out that the following exact relations hold:
\begin{align}
\langle p,s| S_{q}^{i} |p,s\rangle = -\langle p,s| \mathcal{S}^{i}_{1} |p,s\rangle \label{eRel_1}\\
\langle p,s| S_{g}^{i} |p,s\rangle = -\langle p,s| \mathcal{S}^{i}_{2} |p,s\rangle \label{eRel_2}
\end{align} 
This means that regardless of whether these terms vanish or not, the proton matrix elements for $\mathcal{S}^{i}_{1}$ and $\mathcal{S}^{i}_{2}$ will actually cancel the corresponding matrix elements for the `spin' operators $S_{q}^{i}$ and $S_{g}^{i}$ in the angular momentum sum rule. The physical interpretation of the sum rule is therefore lost, and so $L_{q}$/$L_{g}$ can no longer be interpreted as orbital angular momentum observables. It is also clear that the matrix elements $\Sigma$ and $S_{g}$ are not constrained since they do not contribute to the sum rule. A similar cancellation to equation~\ref{eRel_1} was also found in~\cite{shore_white}, although this approach relied on a wave-packet and form factor formulation which was later shown by~\cite{Bakker} to not hold in general. \\

\noindent
The analysis in this section demonstrates that neither conditions 1 nor 2 are satisfied, and therefore the validity of the derivation of the angular momentum sum rule is undermined. Physically speaking, it is the non-perturbative structure of QCD which prevents one from forming distinct quark and gluon observables in this way. These results also provide a resolution to the spin crisis since the cancellation of the $\Sigma$ term in the sum rule lifts the constraint on $\Sigma$, which means that there is no longer an a priori expectation as to what value this matrix element should take.

\section{Conclusions}
\label{section5}

Spatial boundary term operators play an important role in quantum field theories, and in particular the issue of whether they vanish or not is a recurring theme in many of the applications of these theories. The main aim of this paper was to use an axiomatic field theory approach in order to establish a concrete condition on when these terms vanish. It turns out that a necessary and sufficient condition for this class of operators to vanish is that the operator must annihilate the vacuum state. This is a somewhat surprising result in itself because it is completely independent of how this operator acts on the full space of states. In the remainder of this paper we applied this result in order to address the issue of whether meaningful quark-gluon angular momentum operator decompositions are possible in QCD. It turns out that a common feature of these decompositions is the necessity to drop certain spatial boundary terms called \textit{superpotentials}. Using the boundary term conditions established in the previous part of the paper, we analysed the superpotential terms for the specific case of the \textit{Jaffe-Manohar decomposition} derived from the Belinfante energy-momentum tensor, with the conclusion that the sum of these superpotential operators is non-vanishing. In this context, this suggests that the Jaffe-Manohar angular momentum operator decomposition does not hold. An important consequence of these non-trivial boundary operators is the effect that they have on the \textit{proton angular momentum sum rule}. By keeping these boundary terms explicit, we found that the sum rule is modified in a rather surprising way -- the supposed gluon $S_{g}$ and quark $\Sigma$ spin terms are completely cancelled in the expression. This throws into doubt the physical interpretation of these terms and also provides a resolution to the proton spin crisis, since the cancellation of the $\Sigma$ term in the sum rule lifts the constraint on $\Sigma$, and therefore one loses any expectation on what value it should take. Physically speaking, the boundary term conditions imply that the non-perturbative structure of QCD prevents the possibility of forming distinct quark and gluon observables in this way.

\section*{Acknowledgements}
I would like to thank Thomas Gehrmann and J\"urg Fr\"ohlich for useful discussions and input. This work was supported by the Swiss National Science Foundation (SNF) under contract CRSII2\_141847.

\appendix

\section{}
\label{appendix_a}

Because of the way the Belinfante energy-momentum tensor is defined in equation~\ref{belinfante_T}, the Belinfante $P^{\mu}$ and canonical $P_{c}^{\mu}$ momentum operators are related to one another as follows:
\begin{align}
P^{\mu} &= P_{c}^{\mu} + \int d^{3}x \ \partial_{i}B^{i\mu}
\label{2Ps}
\end{align}
where $B^{i\mu} = \tfrac{1}{2}\left(S^{i 0\mu} + S^{0\mu i} + S^{\mu 0 i}\right)$. So $P^{\mu}$ and $P_{c}^{\mu}$ differ by a spatial boundary term. Since $P_{c}^{\mu}$ is the generator of translations this means that: $[iP_{c}^{\mu},F(y)] = \partial^{\mu}F(y)$, where $F(y)$ is any local field. However, because of the relation in equation~\ref{2Ps}, it follows that:
\begin{align*}
[iP^{\mu},F(y)] &= [iP_{c}^{\mu},F(y)] + i\int d^{3}x \ [\partial_{i}B^{i\mu},F(y)] \\
&= [iP_{c}^{\mu},F(y)] + i\lim_{R \rightarrow \infty} \int d^{4}x \ \alpha(x_{0})f_{R}({\bf{x}})\left[\partial_{i}B^{i\mu}(x),F(y) \right] \\
&= [iP_{c}^{\mu},F(y)] -\underbrace{i\lim_{R \rightarrow \infty} \int d^{4}x \ \alpha(x_{0})\left(\partial_{i}f_{R}({\bf{x}})\right)\left[B^{i\mu}(x),F(y) \right]}_{=0} \\
&=\partial^{\mu}F(y)
\end{align*} 
where the second term in the penultimate line vanishes by exactly the same local (anti-)commutativity argument as used in the proof of Theorem~\ref{theorem1} in section~\ref{section2}. So the Belinfante momentum operator $P^{\mu}$ is \textit{also} a generator of space-time translations. Nevertheless, $P^{\mu}$ and $P^{\mu}_{c}$ may well give different results when applied to states in $\mathcal{H}$ since it is not necessarily the case that the spatial boundary operator: $\int d^{3}x \ \partial_{i}B^{i\mu}$ vanishes exactly. The only way to determine this definitively is to apply Theorem~\ref{theorem1}.   \\

\renewcommand*{\cite}{\vspace*{-12mm}}


\begin{thebibliography}{99}

\bibitem{weinberg_1} S. Weinberg, \textit{The Quantum Theory of Fields: Volume 1}, Cambridge University Press (1995). 

\bibitem{maggiore} M. Maggiore, \textit{A Modern Introduction to Quantum Field Theory}, Oxford University Press (2005). 

\bibitem{m&s} F. Mandl and G. Shaw, \textit{Quantum Field Theory}, John Wiley \& Sons (2010).   

\bibitem{greiner} W. Greiner and J. Reinhardt, \textit{Field Quantization}, Springer-Verlag (1996). 

\bibitem{srednicki} M. Srednicki, \textit{Quantum Field Theory}, Cambridge University Press (2007). 

\bibitem{leader_review} E. Leader and C. Lorc\'e, ``The angular momentum controversy: What's it all about and does it matter?,'' \textit{Phys. Rept.} {\bf{541}}, 163 (2014).

\bibitem{weinberg_2} S. Weinberg, \textit{The Quantum Theory of Fields: Volume 2}, Cambridge University Press (1996). 

\bibitem{wightman} R. F. Streater and A. S. Wightman, \textit{PCT, Spin and Statistics, and all that}, W. A. Benjamin, Inc. (1964).

\bibitem{haag} R. Haag, \textit{Local Quantum Physics}, Springer-Verlag (1996). 

\bibitem{ojima_book} N. Nakanishi and I. Ojima, \textit{Covariant Operator Formalism of Gauge Theories and Quantum Gravity}, World Scientific Publishing Co. Pte. Ltd (1990).

\bibitem{strocchi} F. Strocchi, \textit{An Introduction to Non-Perturbative Foundations of Quantum Field Theory}, Oxford University Press (2013). 

\bibitem{ojima_ref1} T. Kugo and I. Ojima, ``Local Covariant Operator Formalism of Non-Abelian Gauge Theories and Quark Confinement Problem,'' \textit{Prog. Theor. Phys. Suppl.} {\bf{66}} (1979).

\bibitem{kastler} D. Kastler, D. W. Robinson and J. A. Swieca, ``Conserved Currents and Associated Symmetries; Goldstone's Theorem,'' \textit{Commun. Math. Phys.} {\bf{2}}, 108 (1966).

\bibitem{strocchi_77} R. Ferrari, L. E. Picasso and F. Strocchi,  ``Local operators and charged states in quantum electrodynamics,'' \textit{Il Nuovo Cim.} {\bf{39A}}, 1 (1977). 

\bibitem{ojima_79} I. Ojima and H. Hata, ``Observables and quark confinement in the covariant canonical formalism of Yang-Mills theory II,'' \textit{Zeits. f. Physik C, Particles and Fields} {\bf{1}}, 405 (1979). 

\bibitem{strocchi_03} G. Morchio and F. Strocchi, ``Charge density and electric charge in quantum electrodynamics,'' \textit{J. Math. Phys.} {\bf{44}}, 5569 (2003).  

\bibitem{emc} M. J. Ashman et al., ``An investigation of the spin structure of the proton in deep inelastic scattering of polarised muons on polarised protons,'' \textit{Nucl. Phys.} {\bf{B328}}, 1 (1989).

\bibitem{belinfante} F. J. Belinfante, ``On the Current and the Density of the Electric Charge, the Energy, the Linear Momentum and the Angular Momentum of Arbitrary Fields,'' \textit{Physica} {\bf{7}}, 449 (1940).

\bibitem{Jaffe_Manohar} R. L. Jaffe and A. Manohar, ``The $g_{1}$ problem,'' \textit{Nucl. Phys.} {\bf{B337}}, 509 (1990). 

\bibitem{Wakamatsu} M. Wakamatsu, ``Is gauge-invariant complete decomposition of the nucleon spin possible?,'' \textit{Int. J. Mod. Phys. A} {\bf{29}}, 1430012 (2014). 

\bibitem{Ji_initial} X. Ji, ``Gauge-Invariant Decomposition of Nucleon Spin,'' \textit{Phys. Rev. Lett} {\bf{78}}, 610 (1997).

\bibitem{jaffe_delta_g} R. L. Jaffe, ``Gluon spin in the nucleon,'' \textit{Phys. Lett} {\bf{B365}}, 359 (1996).

\bibitem{Ji_Tang_Hoodbhoy} X. Ji, J. Tang and P. Hoodbhoy, ``Spin Structure of the Nucleon in the Asymptotic Limit,'' \textit{Phys. Rev. Lett.} {\bf{76}}, 740 (1996).

\bibitem{ji_renorm} X. Ji, ``Breakup of hadron masses and the energy-momentum tensor of QCD,'' \textit{Phys. Rev.} {\bf{D52}}, 271 (1995).

\bibitem{ji_lorentz} X. Ji, ``Lorentz symmetry and the internal structure of the nucleon,'' \textit{Phys. Rev.} {\bf{D58}}, 056003 (1998).

\bibitem{v-axial_paper} J. Pasupathy and R. K. Singh, ``Axial Vector Current Matrix Elements and QCD Sum Rules,'' \textit{Int. J. Mod. Phys. A} {\bf{21}}, 5099 (2006).

\bibitem{lattice_A_sq} O. Pene, B. Blossier, Ph. Boucaud, A. Le Yaouanc, J. P. Leroy, J. Micheli, M. Brinet, M. Gravina, F. De Soto, Z. Liu, V. Morenas, K. Petrov and J. Rodriguez-Quintero, ``Vacuum expectation value of $A^{2}$ from LQCD,'' \textit{Proceedings of Science, The many faces of QCD Conference}, 010 (2010).

\bibitem{pion_form} S. J. Brodsky, C. D. Roberts, R. Shrock, and P. C. Tandy, ``Essence of the vacuum quark condensate,'' \textit{Phys. Rev.} {\bf{C82}}, 022201(R) (2010). 

\bibitem{shore_white} G. M. Shore and B. E. White, ``The gauge-invariant angular momentum sum-rule for the proton,'' \textit{Nucl. Phys.} {\bf{B581}}, 409 (2000).

\bibitem{Bakker} B. L. G. Bakker, E. Leader and T. L. Trueman, ``Critique of the angular momentum sum rules and a new angular momentum sum rule,'' \textit{Phys. Rev.} {\bf{D70}}, 114001 (2004). 


\end{thebibliography}
\end{document}